\documentclass{aa}  
\usepackage{graphicx}
\usepackage{txfonts}
\usepackage{aalongtable}
\usepackage{longtable}

\usepackage{natbib}
\bibpunct{(}{)}{;}{a}{}{,}

\input epsf.sty

\begin{document}
   \title{X-ray observations of the hot phase in Sgr~A*}

   \author{ A. R\'o\.za\'nska \inst{1} 
          P. Mr\'oz \inst{2}
          M. Mo\'scibrodzka \inst{3}
          M. Sobolewska \inst{1}
          T. P. Adhikari \inst{1} }
   \offprints{A. R\'o\.za\'nska}

   \institute{  N. Copernicus Astronomical Center,
		Bartycka 18, 00-716 Warsaw, Poland \\
		\email{agata@camk.edu.pl}
                \and 
                Warsaw University Observatory, 
                Al. Ujazdowskie 4,
                00-478 Warsaw, Poland \\
                \email{pmroz@camk.edu.pl}
                \and
                Department of Astrophysics/IMAPP,
                Radboud University Nijmegen,P.O. Box 9010, 
                6500 GL Nijmegen, The Netherlands  \\
                M.Moscibrodzka@astro.ru.nl
                }

   \date{Received ????, 2007; accepted ???, 2007}



\abstract
   {We analyze 134 ks Chandra ACIS-I observations of the Galactic Centre (GC) 
   performed in July 2011. The X-ray image with the field of view $17' \times 17'$ 
  contains the hot plasma surrounding the Sgr~A*. The obtained surface brightness map
  allow us  to fit Bondi hot accretion flow to the innermost hot plasma around the GC.
  }
   { Contrary to the  XMM-Newton data where strong 6.4 keV iron line was observed and
interpreted as a reflection from molecular clouds, we
   search here for the diffuse X-ray emission with prominent 6.69 keV iron line.
The surface brightness profile found by us allowed to determine the stagnation radius 
of the flow around Sgr~A*.
}
   {We have fitted spectra from region up to $5''$ from Sgr~A* using a thermal 
   bremsstrahlung  model and four Gaussian profiles responsible for K$_{\alpha}$ 
  emission lines of Fe, S, Ar, and Ca. The 
  X-ray surface brightness profile up to $3''$ from Sgr~A* found in our data image, 
  was successfully  fitted with the dynamical model of Bondi spherical accretion.
  }
   {We show that the temperature of the hot plasma derived from our spectral 
   fitting  is of the order of 2.2-2.7 keV depending on the choice of background. 
   By modelling the surface brightness profile, 
   we derived the temperature and number density profiles in the vicinity of the black hole.
   The best fitted model of spherical Bondi accretion shows that this type of flow 
  works only up to $3''$ and implies outer plasma density and temperature to be:
 $n_{\rm e}^{\rm out}=18.3 \pm {0.1}$ cm$^{-3}$ and $T_{\rm e}^{\rm out}= 3.5 \pm {0.3}$ 
 keV respectively.}
   {We show that the Bondi flow can reproduce observed surface brightness profile 
 up to $3''$ from Sgr~A* in the Galactic Center.  This result strongly suggests the position of 
 stagnation radius in the complicated dynamics around GC.
The temperature at the outer radius of the flow is higher by 1 keV,  than this found 
by our spectral fitting of thermal plasma within the circle of $5''$. The Faraday rotation computed
from our model towards the pulsar PSR J1745-2900  near the GC agrees with the observed one, 
recently reported. 
We speculate here, that the  emission lines observed in spectra up to $5''$   
can be interpreted as  the reflection of the radiation from two-phase regions
   occurring at distances between $3-5''$ of the flow. The hot plasma in Sgr~A* illuminated 
  in the past by strong radiation field may be a seed for 
   thermal instabilities and eventual strong clumpiness.
   }

\maketitle
%

\section{Introduction}
\label{sec:src}

The central region of the Milky Way hosts 
a supermassive black hole (BH) with well established mass 
$M_{\rm BH}=4.4 \times 10^6 M_{\odot}$ \citep{genzel2010},
 at a distance $R=8.4$ kpc. 
The surroundings of the black hole are unique. 
It contains a number of relatively cold molecular 
clouds \citep{zhao2009} located within $15'$ from Sgr~A*, and a hot plasma 
of temperature around 2 keV situated within a few arcmin \citep{baganof2003}. 
Both phases of gas show X-ray emission with prominent iron K$_\alpha$ line,
which may arise due to reflection of hard radiation emitted at 
the very center of the Galaxy nucleus.

The cold phase emits neutral Fe K$_{\alpha}$ line at energy 6.4 keV, 
as in case of molecular clouds reported recently by \citet{ponti10}. 
These cold clouds situated at the distance of 30-60 pc from the Galactic Centre (GC), 
are well resolved by the {\it XMM-Newton} satellite, enabling proper 
studies of their velocities and the nature of the reflected continuum.
Tracing history of fluorescent reflection, it is possible to capture
historic luminosity variations of Sgr~A* \citep{yunwei2011}. 

The hot phase emits  the Fe K$_{\alpha}$ line at energy 6.690 keV, 
indicating that iron atoms are in  helium-like ionization state.
 X~-~ray emission from a diffuse hot gas was observed in extended regions 
around the GC \citep{park2004,muno2004}. Particularly, it appears to be concentrated 
within the central 2-3 pc eastwards from Sgr~A*, named Sgr~A~East, and most 
probably is caused by a young supernova remnant (SNR) \citep{maeda2002}. 
Additionally, the hot phase lies on the line of sight toward Sgr~A* and 
it may accrete on supermassive black hole in the form of 
spherical Bondi accretion or radiatively inefficient accretion flow (RIAF).
The dynamical model of the accretion on the GC is still not well established 
since it is clearly seen that this region may be a mixture of
hot plasma with cold clouds i.e. mini-spiral region \citep{zhao2010}.
The accretion of hot plasma  fed by stars due to mass and energy exchange
was proposed by \citet{shcherbakov2010}
to explain the observed surface brightness profile. On the other hand,  
\citet{czerny13} constructed a model of multiple accretion events caused by cold 
clouds to trigger the Sgr~A* activity.  Such hot and cold two-phase medium
can be formed due to thermal instability caused by irradiation with hard X-rays 
 from the GC, as recently shown by \citet{rozanska14}.
Thermally unstable clouds can be located close to the center 
at 0.008-0.2 pc, i.e. within $5''$,  and they can survive hundreds years.
The radiation reflected from those clouds can produce hot iron line 
in that region.

 Since the extension of the hot gas is an order of magnitude lower
than of the cold phase, {\it Chandra} and {\it Suzaku}
satellites are better for detailed studies of those regions
\citep[e.g.][and references therein]{maeda2002,koyama2007,wang2013}.
 Recently, \citet{wang2013} have reported results of spectral fitting of the longest 3 Msec
{\it Chandra} observations of the $1.5''$ circular region around Sgr~A*, and the spectrum 
presented in this paper is the most detailed ever published. 
But the surface brightness profile from dynamical model is not fitted in this paper. The
authors fit model spectrum computed as a thermal emission from the distribution of temperatures and densities according to the analytical formula appropriate for RIAF (see Sec.~\ref{sec:wang} in this paper.

The purpose of this paper is to study the hot plasma around Sgr~A*.
Analysing X-ray image we aim to put constraints on the physical 
parameters of the hot phase, which may be an important material 
to be a source of radiation from Sgr~A*. 
We present archival {\it Chandra} ACIS-I observations
of the GC made in July 2011 with the field of view $17' \times 17'$ containing 
both the Sgr~A* and Sgr~A~East. The data were collected with three time intervals, 
very short one after another. 
Therefore, spectral and imaging coadding was very accurate. 

From {\it Chandra} observations, 
we extracted spectrum from one circular region
around Sgr~A* with radius of  $5''$. We performed detailed 
spectral analysis by fitting  a thermal 
plasma with a strong  K$_\alpha$ line from helium-like iron.
Additionally, we fitted other emission lines seen in residuals form C, Ar, Ca, 
but the detection of last two was marginal.
Following \citet{shcherbakov2010}, we extracted surface brightness profile 
around Sgr~A* within the radius of 
$10''$ with sub-pixel accuracy.  We fitted the dynamical model of the hot Bondi accretion 
flow by \citet{moscibrodzka2009} to this imaged emissivity profile.
To make a fit,  for each computed model we constructed the map of emission around GC 
convolved with {\it Chandra} exposure map of ACIS-I chip. 
In our observations  Sgr~A* is located off-axies, and 
we constructed point spread function (PSF) to account for this particular 
observation. Final models were convolved with normalized PSF, before 
comparing to the data. 
We point out here, that only basic spectral fitting is done in this paper
due to the short time exposure. In contrast to \citet{wang2013},
we do not compute the line emissivity from our model. Instead, we focus on modeling 
the sub-pixel surface brightness profile \citep[see also][]{shcherbakov2010}.

The main result of the observed surface brightness profile modelling 
is the derivation of electron density and temperature profiles 
of the flow from $3''$ down to the black hole horizon. 
The best fitting model is for temperature and density at the 
outer flow radius  $T_{\rm e}^{\rm out} =3.5 \pm 0.3 $ keV, and 
$n_{\rm e}^{\rm out}=18.3 \pm 0.1 $ cm$^{-3}$.
Our fit is not valid outside $3''$ indicating the 
location of the stagnation radius
of the dynamical flow around GC. Ouside stagnation radius, 
matter can possess outflow as it was indicated in RIAF model of \citet{wang2013}.
The temperature at the outer flow radius 
is in good agreement with that derived from spatially resolved 
spectral modelling presented in \citet{wang2013}, and then rejected by RIAF model. 
In addition, \citet{wang2013}  considered RIAF model, where  
the line emissivity was computed from the predicted RIAF temperature profile. 
Such line fitted to the data indicated the outer plasma temperature to be 1 keV, 
which is lower than in our model.
In this paper we compare temperature and density profiles implied for 
both models, RIAF and Bondi.

Fitting the canonical model of thermal bremsstrahlung plus Gaussian profiles for all lines
indicates temperatures for the plasma around Sgr~A* in the 2.2--2.7\,keV range, 
depending on the choice of background.  And again, those values 
are lower than that obtained by \citet{wang2013} with the same model for 
the $1.5''$ Sgr A* spectral fitting, but higher then 1 keV temperature 
derived by them using RIAF emissivity profile.
The EW of the iron line we find is in the $0.9-1.2$\,keV range.
Within the uncertainties, our EW of the iron line is
consistent with that estimated by \citet{wang2013} at the $1\sigma$ confidence level.
Since the result is degenerated according to plasma temperature and warm absorption,  
the broad-band spectra are needed to discriminate between models.

In addition, we calculated the Faraday rotation (RM) towards the Galactic Center and 
towards the pulsar located at $3''$ in the projected distance away from the GC. 
The first result matches with observations \citep{marrone2007}
only with additional assumption of the very weak magnetic field, while the second 
result agrees with the recently observed value very well \citep{eatough2013}. 

Our paper shows that Bondi accretion can well represent the hot phase accretion
on the GC up to $3''$.  We argue that the radiation originating from the 
very close vicinity of Sgr~A*, illuminating external zones of the Bondi flow
may produce two-phase medium when all lines are created. The exact distance
of such clumps and their lifetime are given in \citet{rozanska14}.
 The temperature of such clouds varies from log$(T)= 5.5 -7.5$ in Kelvins, and 
the warm clouds have longer lifetimes than the cold clouds. We postulate here, that 
they  can be 
responsible for the formation of iron line at energy 6.7 keV, seen in the 
{\it Chandra} data. Better data are needed to fully confirm this hypothesis.

The structure of the paper is as follows: in Sec.~\ref{sec:obs} we present the  
details of data extraction that resulted in obtaining  images, brightness profile
and spectra of hot plasma in the neighbourhood of the GC.
Sec.~\ref{sec:fit} contains results of spectral fitting, 
while Sec.~\ref{sec:flow} shows results of surface brightness 
profile fitting in the vicinity of Sgr~A*. 
Discussion and conclusions are presented in Sec.~\ref{sec:summary}.

\section{Observations}
\label{sec:obs}

The {\it Chandra} X-Ray Observatory has observed the central regions of 
the Galaxy multiple times. For our analysis, we have selected three observations
carried out in July 2011. the longest unpublished observations available in the 
{\it Chandra} archive. They were performed for a very similar period of time, 
allowing to avoid significant changes with time and astrometry. Table~\ref{tab:obs} 
presents  identification numbers, starting dates and exposure times of the observations.  
All of them were done using ACIS-I spectrometer with the field of view 
17' $\times$ 17'.
It corresponds to the region of 40 $\times$ 40 pc in the center of our Galaxy.
The pointing coordinates were 
$\alpha=17^{\textrm{h}}45^{\textrm{m}}59.^{\textrm{s}}70$, 
$\delta=-28^{\circ}58'15.''9$, and were $4.'8$ away from Sgr~A*.

\begin{table}
\begin{center}
\caption{{\it Chandra} satellite observation log.}
\label{tab:obs}
\begin{tabular}{cccc}
\hline 
ObsID & Date & Exp. time & GTI \\
   &      & [ks] & [ks] \\ \hline
12949 & 2011-07-21 & 58.48 & 48.88 \\
13438 & 2011-07-29 & 66.18 & 59.79 \\
13508 & 2011-07-19 & 31.49 & 26.10 \\ \hline 
\end{tabular}
\end{center}
\end{table}


\begin{figure}
\hspace{-0.8cm}
\begin{tabular}{c} 
\includegraphics[angle=0,width=9.5cm]{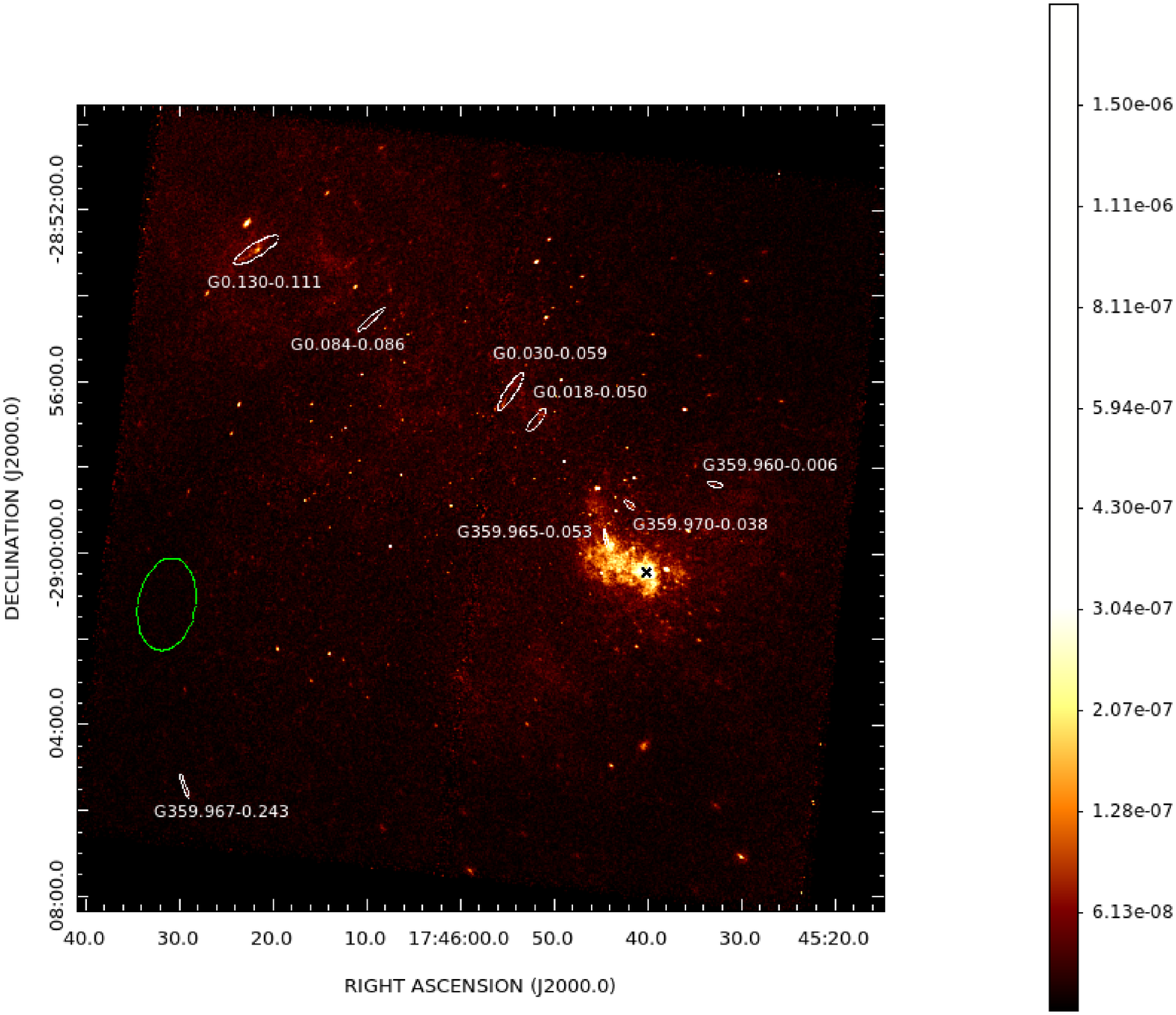} \\
\includegraphics[angle=0,width=9.5cm]{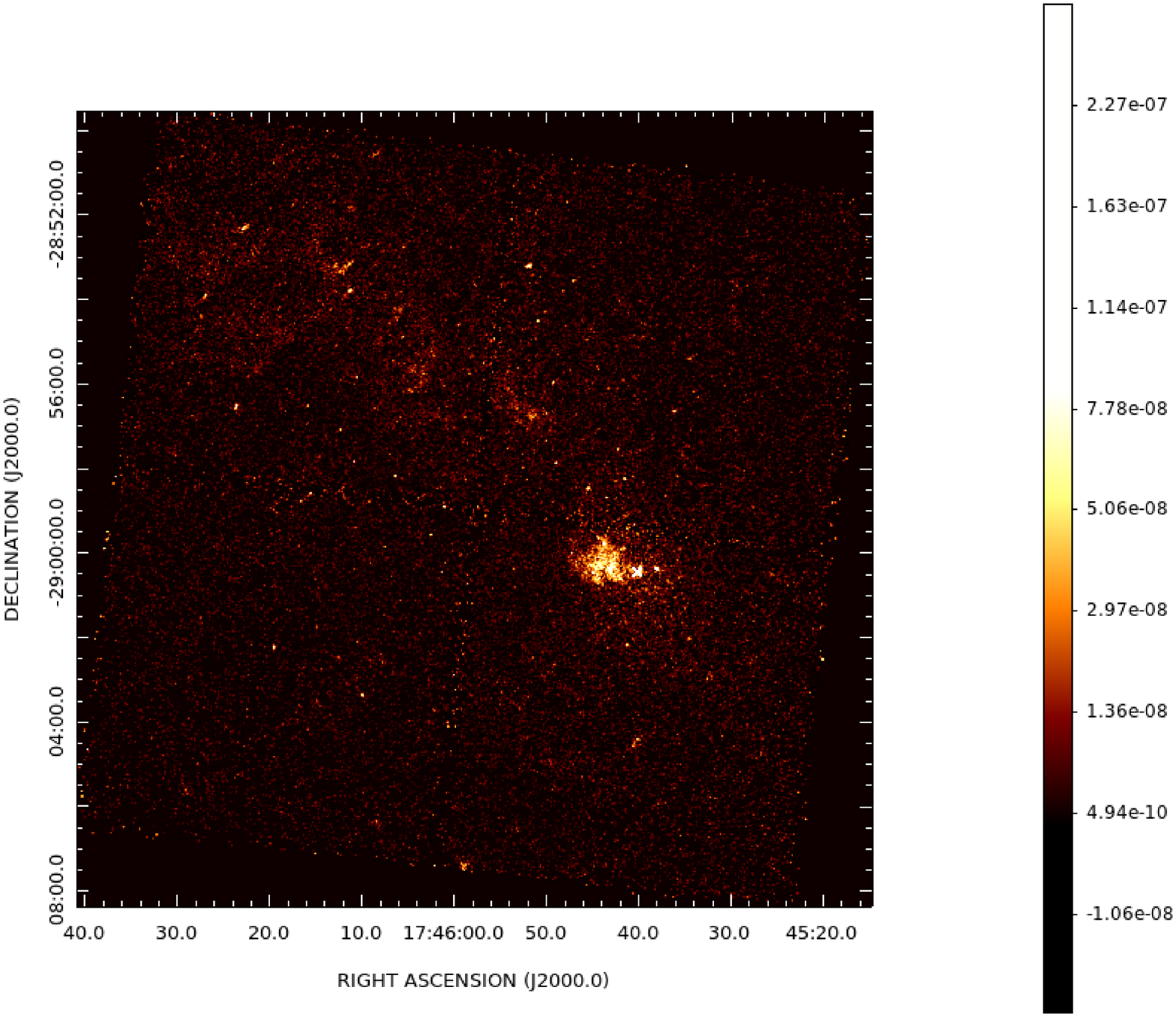}
\end{tabular}
\caption{Flux image of the GC in the 0.5-8 keV energy range (upper panel). Sgr~A* is marked with a black 
cross. A few brightest filaments reported by \citet{johnson09} are marked with white
(small) ellipses and named following the names given in previous detection . 
The green (large) ellipse shows the region used for background spectrum extraction (the case of distant
background, see Sec.~\ref{sec:ext} for details).
Bottom panel shows flux image of the GC in the energy range of 6.3-7.0 keV, where iron line 
is created. Emission presented here is the K$_{\alpha}$ line emission after subtracting the continuum flux.
The flux is presented in asinh scale, since it behaves well for negative count rate values
which might originate during subtraction process.}
\label{fig:flux}
\end{figure}


\subsection{Data reduction}
\label{sec:cloudy}


All subsequent analysis was performed using CIAO 4.4 
software\footnote{http://cxc.harvard.edu/ciao/index.html}, and CALDB version 4.5.3.

We started our analysis by processing  individual observations.
The corrections for charge transfer inefficiency, bad pixel 
removal, and background flaring were done in the whole energy band, 0.5-8.0 keV. 
The {\sc lc\_sigma\_clip}
 tool was used to clip the  data that deviates from the mean by more than 2 sigma. 
In order to minimize the influence of the molecular clouds variability
we made two tests: 
i) point sources were removed from the ACIS-I2 CCD chip, where diffuse 
emission was marginal, filtering counts to the 0.5-8 keV energy band  
ii) photons with energy from 7 to 10 keV in the whole ACIS-I field of view were considered. 
In both cases similar good time intervals (GTI) were obtained. 
Hereinafter, we decided to use 
GTI found in case (i), as they set more stringent limits. For all 
observations the light curves are of good quality and the background flaring 
is insignificant. Finally, we choose GTI=134.77 ks, which covers 86\% of the total 
exposure time.

Although during all observations the telescope was pointed at the same target, 
we have found a small shift between three images (a few arcsec). In order to 
correct the absolute astrometry of the data, we searched for point sources 
using the {\sc wavdetect} tool. 53 bright sources were found and their coordinates 
were obtained from the SIMBAD astronomical database
\footnote{http://simbad.u-strasbg.fr/}. Subsequently we have applied minor corrections 
to the WCS using the {\sc reproject\_aspect} script, which 
calculates shift between the {\sc wavdetect} output and the absolute 
coordinates from a catalog. 

\subsection{Images in continuum and in K$_\alpha$ iron line bandpass}

We have regraded event files to a common tangent point with 
{\sc reproject\_events} 
and we have merged them using {\sc dmmerge}. Similar procedure was applied 
to the images, beside that the {\sc reproject\_image} and {\sc dmimgcalc} 
scripts were used instead.  
We have created exposure-corrected image in the 0.5-8 keV energy band using 
the {\sc fluximage} script. The flux image is presented in Fig.~\ref{fig:flux} upper panel,
with the position of Sgr~A* marked by the black cross. At this figure we 
have also indicated and named the brightest filaments reported by \citet{johnson09}. 
Nevertheless, our total exposure time was too short to make spectral analysis  of these
filaments or discover new ones, so we do not proceed with their analysis.
However, by merging the three observations we obtained enough counts to  
study the hot extended plasma around  Sgr~A*. Since 
{\it Chandra} X-ray observatory has the best spatial resolution among all 
currently working satellites, in Sec.~\ref{sec:flow} we present analysis of the surface 
brightness profile around Sgr~A* with $1''$ scale resolution. 

Since we were interested 
in the spatial distribution of the ionized iron K$_{\alpha}$ line, additional 
images were created in the 4.5-6.3 keV (the continuum) and 6.3-7.0 keV (broad line) 
energy bands. The former was subtracted from the line image with a normalization factor
of 0.12. This constant was calculated on the assumption that the continuum 
emission is modelled by an absorbed power-law function with photon index $\Gamma=3$ 
and hydrogen column density $N_{\rm H}=10^{23}\ \textrm{cm}^{-2}$. The continuum 
spectrum was modeled using Sherpa\footnote{http://cxc.harvard.edu/sherpa4.4/} 
\citep{freeman2001} and simulated 
with the {\sc fake\_pha} command. 
The response matrix and effective area files were created 
with {\sc mkacisrmf} and {\sc mkwarf} (sample RMFs and ARFs were also taken 
from CALDB, similar value was obtained). 
The normalization factor is simply the ratio of the flux in the line to the flux in
 the continuum energy band. 

The image in iron line emission after subtracting 
continuum image is presented in the bottom panel of Fig.~\ref{fig:flux}. The majority of point sources 
and diffuse 
emission around Sgr~A* are properly subtracted.  Some faint molecular 
clouds remained after subtraction because the emission from the cold 
Fe line overlaps broad line energy band \citep{ponti10}.
We have constructed the flux image in K$_{\beta}$ line, but the emission was negligible. 

Fig.~\ref{fig:flux} bottom panel is presented in the asinh
(the inverse hyperbolic sine) scale. In case of the high count rate this scale
is practically logarithmic; for low values it changes linearly. It is 
especially useful in subtracted images i.e. the iron line, as it 
 behaves well for negative count rate values which might be present.

\subsection{Sgr~A* surface brightness profile}
\label{sec:pro}

 The surface brightness profile can be constructed in 
counts per pixel squared as a function
of distance from the BH. It was previously done for much longer observations
\citep{shcherbakov2010} with the exposure time 953 ks. 

Here, we repeat this analysis for our data, but we fit them with a different
model (see below Sec.~\ref{sec:flow}). 
The size of {\it Chandra} pixel is $0.5''$, but the position of 
the satellite over the duration of observation can be determined with the $0.1''$ 
accuracy by comparing with the known positions of bright point sources. 
As noted by \citet{shcherbakov2010} we can achieve the pixel resolution 
accuracy in the surface brightness profile from knowing the orientation 
of the detector pixels at the given time.

\begin{figure}
\includegraphics[angle=0,width=9cm]{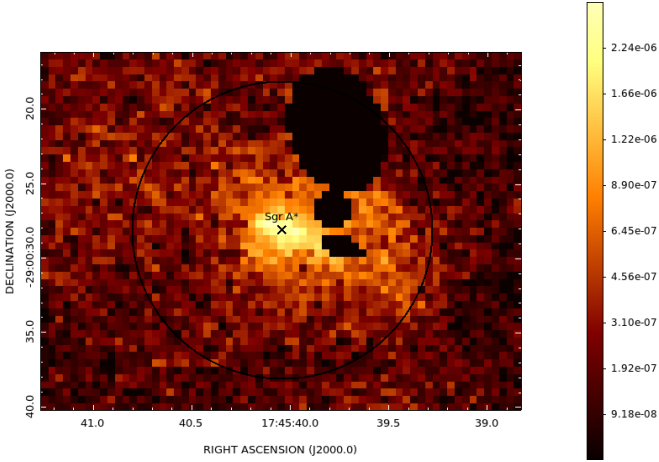}
\caption{X-ray image of the $10''$ region around Sgr~A* (circle). Surface 
brightness profile was extracted integrating counts over annuli from 0 
to $10''$. The influence of point sources was rejected as shown in the image. }
\label{fig:profil}
\end{figure}

\begin{figure}
\hspace{0.2cm}
  \includegraphics[angle=0,width=8.7cm]{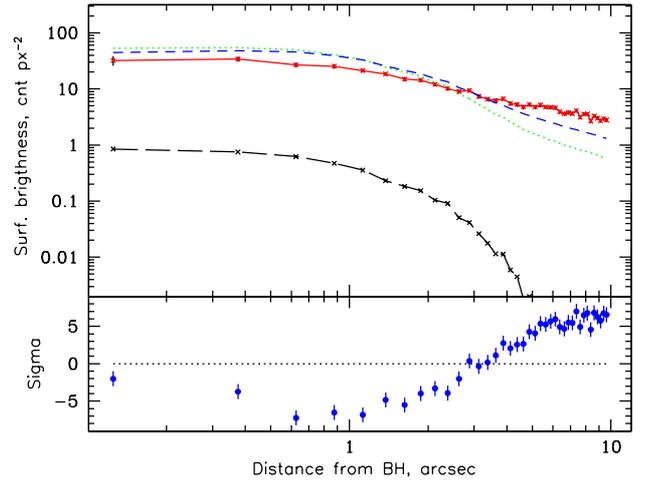}
\caption{Vicinity of the black hole up to $10''$ from Sgr~A*. Red points with 
errors represent observed profile, and are connected by continuous line.
Black long dashed line is 
normalized PSF. Two models convolved with PSF are also presented: green dotted line 
for $T_{\rm e}^{\rm out}=1$ keV, while blue short dashed line for $T_{\rm e}^{\rm out}=3.5$ keV. Residuals are 
shown for best fitted model in the range up to $10''$, i.e. for $T_{\rm e}^{\rm out}=3.5$ keV. }
\label{fig:bright}
\end{figure}

In Fig.~\ref{fig:profil} we show pixel image of the close vicinity 
of the GC.  It was necessary to exclude some point-like 
features in the Sgr~A* neighbourhood because they broke spherical 
symmetry around the source. In order to obtain a radial profile, 
the net counts in a set of concentric annuli around Sgr~A* were measured 
with {\sc dmextract} tool. The width of annuli is $0.25''$, 
the radius of the outer annulus is $10''$, and is marked in 
Fig.~\ref{fig:profil} by a black circle. 

The radial profile is monotonically decreasing in the whole range of radii. 
The final profile of the surface brightness with errors is presented in Fig.~\ref{fig:bright} 
by red points. 

To make any further analysis of the brightness profile on the sub-pixel scale,
we have to account for the point spread function (PSF) appropriate for 
the {\it Chandra} X-ray telescope. Any model fitted to the profile has to be 
convolved with the PSF.  
To construct the ACIS-I PSF for our observation we used 
the {\it Chandra} ray-tracing program,
ChaRT\footnote{http://cxc.harvard.edu/chart/runchart.html},
to simulate the Sgr~A* (point source) photons scattered by the {\it Chandra} mirrors,
and the MARX\footnote{http://cxc.harvard.edu/chart/threads/marx/} software
version 4.5.0 to projects the simulated rays onto the detector plane. The MARX
output was used to extract the PSF profile for convolution with our model spectra.
ChaRT requires the position of the point source on the chip, exposure time and
the point source spectrum to run the simulations. For these we used the position
of Sgr~A* in our stacked observation, total GTI exposure, and the best fitting model
to the innermost region of Sgr~A* (5'',  Table~\ref{tab:simpsd}).
The normalized PSF constructed for our particular observation is presented 
in Fig.~\ref{fig:bright} by black crosses and long dashed line.
The PSF is wider than in the paper by \citet{shcherbakov2010}, 
due to the fact that in our observations Sgr~A* was slightly off-axis. 

\subsection{Extraction of the spectrum}
\label{sec:ext}

We  have extracted spectra from regions where the emission of 
the line was the highest, especially from the region 
around Sgr~A*. Sgr~A~East region is treated separately and it will 
be presented in the forthcoming paper.

\begin{table*}
\caption{ Best fit parameters for the $5''$ region around SgrA*. We present two cases modelled
with two different statistics: CSTAT for the case where 
source and background are fitted simultaneously, and $\chi^2$ for background subtracted analysis. 
In the second case we consider two background regions: a distant background marked with the 
green ellipse in the upper panel of Fig.~\ref{fig:flux}, and  a local background defined as an  
annulus between $6'' -18''$ around the $5''$ 
source region. In the first row we show results obtained with a model consisting of  
thermal bremsstrahlung with Galactic absorption (tbabs) and Fe  Gaussian line only. 
The second row shows model parameters when three more 
Gaussian lines are added to account for S, Ar, and Ca emission. 
Both fits were done using CSTAT statistic.  
The third and fourth rows describe background subtracted analysis 
(using the local and distant backgrounds, respectively) performed on data grouped according 
to S/N = 3 using the $\chi^2$ statistic. }
\label{tab:simpsd}
\begin{tabular}{ l c c r l r l r l r l  c c }

\hline

Bkgr. & N$_{\rm H}$ & kT [keV] & Line & E$_{\rm K\alpha}$ [keV] & Line & 
E$_{\rm K\alpha}$ [keV] & Line & E$_{\rm K\alpha}$ [keV] & Line & E$_{\rm K\alpha}$ [keV] 
& Stat.& d.o.f. \\
 & & & & &  & & &  &  &    & &  \\
Reg. & $10^{22}$\,cm$^{-2}$  &  & &  EW [keV]  & & EW [keV] & & EW [keV]  & & EW [keV] &  Val.  &   \\
  & & & & &  & & &  &  &    & &   \\
\hline
\hline
&  7.44$^{+0.15}_{-0.15}$   & 3.57$^{+0.08}_{-0.09}$ & Fe & 6.728$^{+0.035}_{-0.032}$ &
  & ---  & &--- &  &--- &  CSTAT  & \\
Distant  & &  & & & & & &  &  &    & & \\
& & & & $ 0.58 \pm 0.06 $ & & & &  &  &    & 1759 & 1019 \\
 & & & & &  & & &  &  &    & & \\
\hline 
&  9.15$^{+0.27}_{-0.25}$   & 2.66$^{+0.07}_{-0.07}$ & Fe & 6.739$^{+0.064}_{-0.042}$  & 
S & 2.480$^{+0.0299}_{-0.0319}$ & Ar & 3.153$^{+0.05}_{-0.03}$  & Ca &  3.861  
& CSTAT & \\
Distant& & & & &  & & &  &  &    & &  \\
& &   &  &$0.91^{+0.40}_{-0.24}$ & & 0.18$^{+0.04}_{-0.05}$  & & 0.05$^{+0.02}_{-0.02}$&  & 0.03$^{+0.02}_{-0.01}$ & 1254 & 1008 \\
 & & & &  & & & &  &  &    & & \\
\hline 
& 10.52$^{+0.42}_{-0.39}$   & 2.24$^{+0.07}_{-0.07}$ & Fe & 6.745$^{+0.118}_{-0.069}$ & 
S & 2.489$^{+0.030}_{-0.030}$ & Ar & 3.215$^{---}_{---}$ & Ca & 3.865$^{---}_{---}$ 
& $\chi^2$ & \\
Local & &  & & & & & &  &  &    & & \\
 & & & & $1.19^{+0.89}_{-0.48}$ & & 0.20$^{+0.09}_{-0.09}$  & & 0.06$^{+0.04}_{-0.03}$
 & & 0.05$^{+0.04}_{-0.03}$  & 117.32 & 208 \\
& &  & & & & & &  &  &    & & \\
\hline
 &  9.45$^{+0.31}_{-0.27}$   & 2.58$^{+0.07}_{-0.09}$ & Fe & 6.731$^{+0.064}_{-0.045}$  &
S & 2.488$^{+0.012}_{-0.028}$ & Ar & 3.140$^{+0.059}_{-0.040}$ & Ca & 3.866$^{+0.055}_{-0.086}$ 
& $\chi^2$ &  \\
Distant & & &  & & & & &  &  &    & &  \\
& &  &  & $0.92^{+0.32}_{-0.26}$  & & 0.21$^{+0.06}_{-0.06}$ &   &0.06$^{+0.03}_{-0.03}$ &
 & 0.04$^{+0.03}_{-0.02}$ & 122.64 & 198 \\
& & &  & &  & & &  &  &    & & \\
\hline 
\end{tabular}
\end{table*}

 For the purpose of this work, spectrum from circular region of the radius 
of  $5''$ around Sgr~A* was extracted 
and analysed in details. This was done to see
how far from the BH the iron line emission is crucial, and to 
estimate the temperature at the outer radius of the dynamical flow, which 
is independently derived while the surface brightness profile is fitted 
as presented in Sec.~\ref{sec:bondi}.

Prior to creating the spectrum, a few point-like sources were excluded since 
they might contaminate the results. The spectrum was obtained with the 
{\sc specextract} script for each observation independently and merged with 
the {\sc combine\_spectra} tool.
We consider two different regions for background: a distant region 
marked as a green ellipse in the upper panel of Fig.~\ref{fig:flux}, and a local region defined as an
annulus between $6'' -18''$ around the $5''$ source region.  
The distant  background region does not contain any point sources, 
and the diffuse emission is negligible inside the ellipse. In the case of local 
background the possible point source contamination was investigated and corrected for
during the data reduction process. 
After background subtraction, the $5''$ source photon's number 
in the 0.5-8 keV band is equaled to  2937 and 2358, for case distant and local backgrounds, respectively.
Full description of our spectral analysis is presented below in Sec.~\ref{sec:fit}.

\section{Spectral fitting}
\label{sec:fit}

\begin{figure}
\includegraphics[width=85mm]{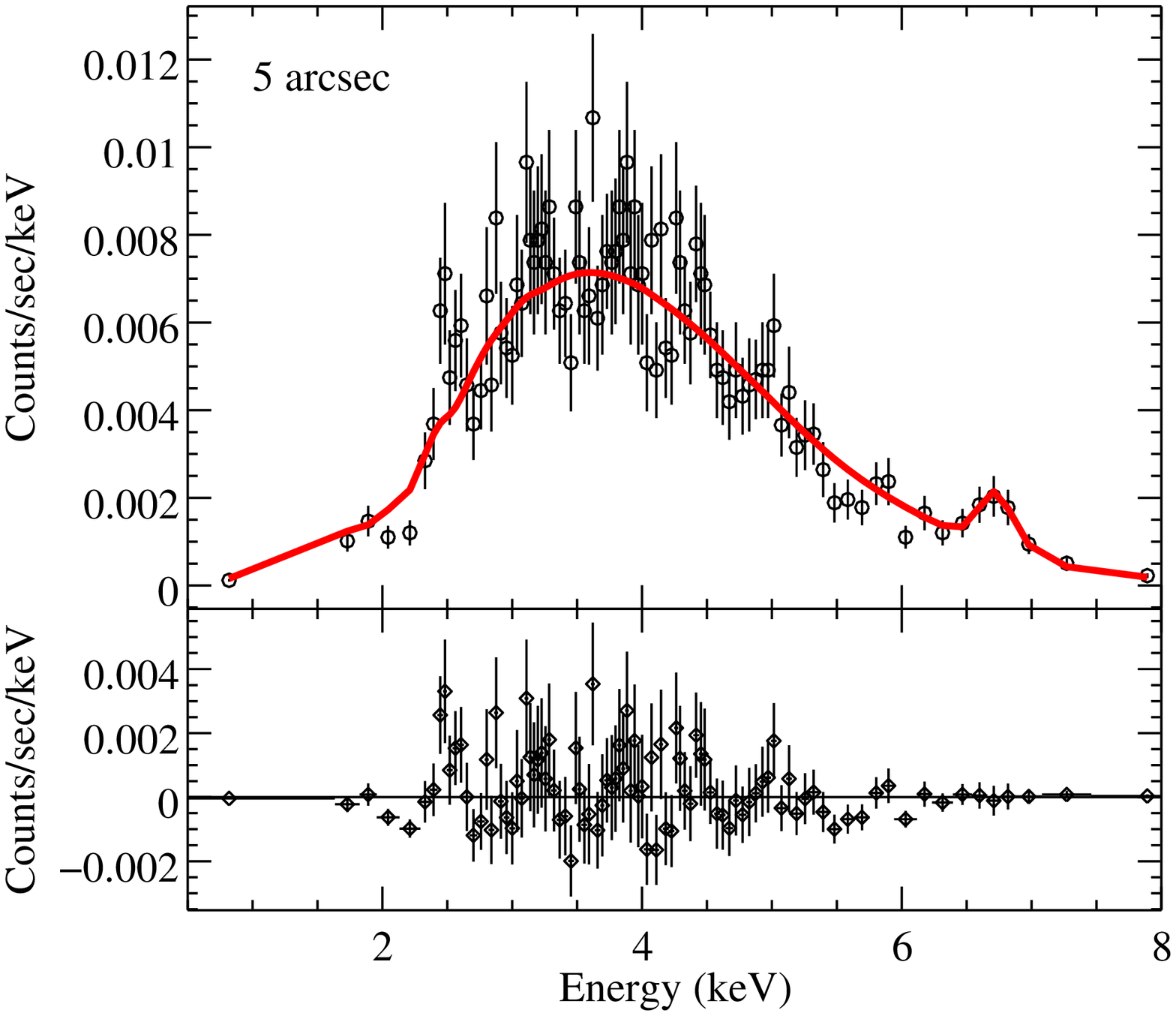}
\includegraphics[width=85mm]{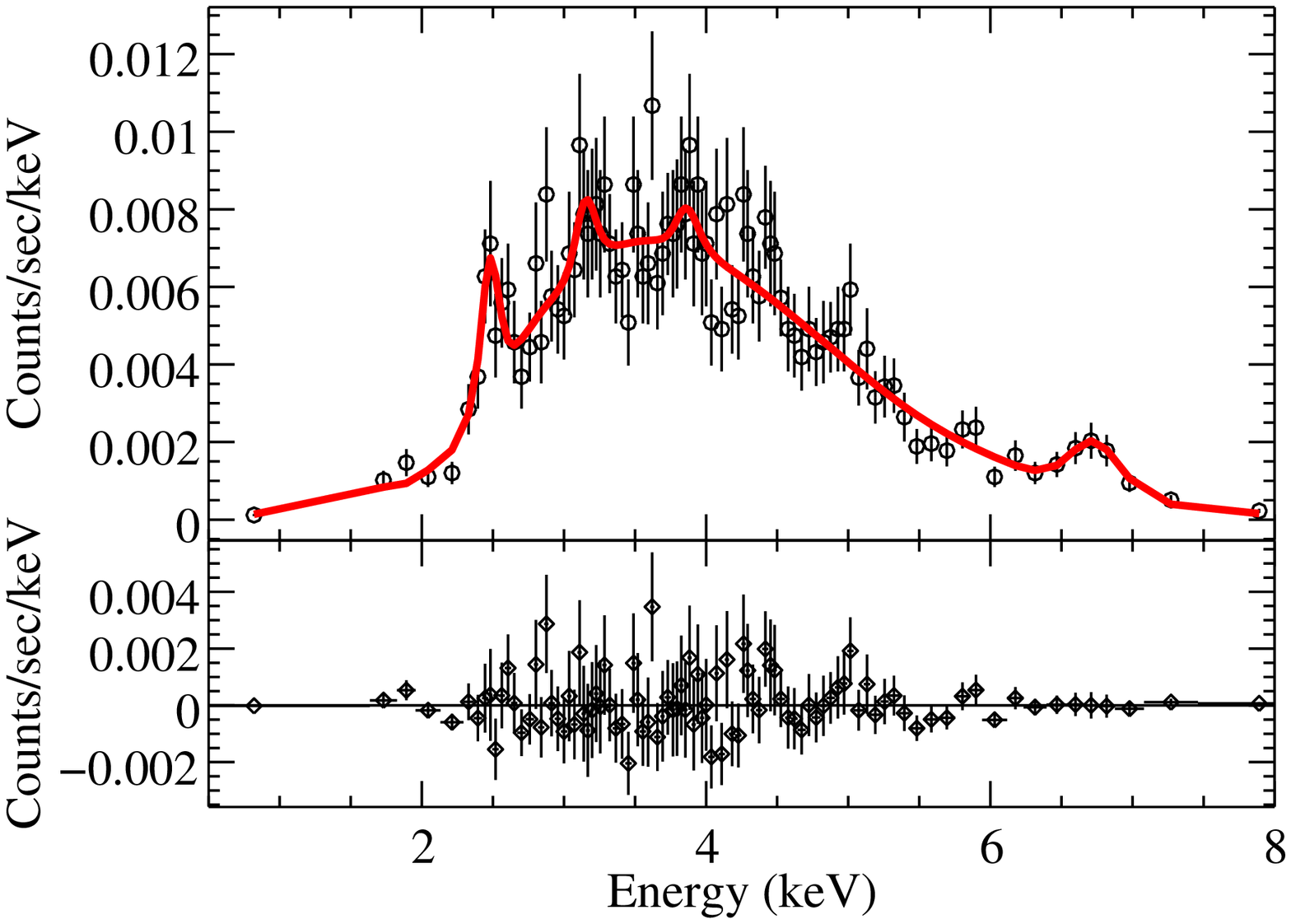}
\caption{Modelling of the spectrum extracted from the $5''$ circular region
around the Sgr~A*. Both panels present fits with distant background.
The upper panel shows a model consisting of Galactic
absorption (tbabs), thermal bremsstrahlung and one Gaussian line to account for the iron emission.
The lower panel shows a model with three additional Gaussian lines for the S, Ar and Ca emission.} 
\label{fig:brems}
\end{figure}


The spectral fitting of the data was performed with the Sherpa 4.4 fitting package.
In all plots presented in this section, the data were 
grouped requiring the signal-to-noise ratio $S/N = 5$, for presentation purposes only. 
All errors on the spectral parameters correspond to the 68\% confidence level
(1$\sigma$) for one significant parameter.

In the first step, considering case A (distant background), we rejected
simple background subtraction process. Instead, the source region and 
corresponding background were modelled simultaneously 
using the CSTAT statistic.
We used a phenomenological model composed of a thermal bremsstrahlung emission 
({\sc bremss}) to describe the  continuum, and a Gaussian profile
to account for the iron K$_{\alpha}$ line emission.  The total model was 
multiplied by warm Galactic absorption ({\sc tbabs}).
The background data were parametrized with a power law model.
Resulting fit parameters are presented in Tab.~\ref{tab:simpsd} in 
the first row. 
Figure~\ref{fig:brems} (upper panel) shows the data and the best fit
model together with residuals. The fit resulted in CSTAT=1759 for 1019 degrees
of freedom (d.o.f.). The residuals below 4\,keV hint at the presence of additional
emission lines detected e.g. by \citet{wang2013}.

 In the second step, using the same source/background regions statistics,
 we add three fit Gaussian profiles to the model to account for S, Ar, and Ca emission lines
reported by \citet{wang2013}. 
The widths of these additional lines were fixed at $\sigma=10^{-5}$\,keV.
Certain residuals around the energy of the S line were visible also in the background data,
and thus the S line was added also to the background model (again with $\sigma$ fixed at $10^{-5}$\,keV).
These changes resulted in ${\rm CSTAT}=1696$ for 1007 d.o.f., a decrease in the temperature of the plasma,
and increase in the column density of the absorber. 

Further improvement to the fit quality was obtained by adjusting the curvature 
of the background model, i.e.
replacing the simple power law model with a 3rd degree polynomial, and 
adding a Gaussian with negative normalization to
account for a feature visible below 1\,keV. These updates did not affect the values of 
plasma temperature nor the column density,
but they led to a substantial decrease in the fit statistics, ${\rm CSTAT}=1254$ for 1011 d.o.f. 
The best fit parameters are shown in the second row of Tab.~\ref{tab:simpsd}. 
The data and best fit model are presented in Fig~\ref{fig:brems}, lower panel.

We investigated if the change in the values of the $N_{\rm H}$ 
(from $7.44^{+0.15}_{-0.15} \times 10^{22}$\,cm$^{-2}$ to $9.15^{+0.27}_{-0.25} \times 10^{22}$\,cm$^{-2}$) 
and $kT$ (from $3.57^{+0.08}_{-0.09}$\,keV to $2.66^{+0.07}_{-0.07}$\,keV)
could be due to a degeneracy between these two parameters. Indeed we found that at each 
step we are able to find both
a low-$N_{\rm H}$ high-$kT$ solution, and a high-$N_{\rm H}$ low-$kT$ solution with 
the statistics that differed only
by $\Delta{\rm CSTAT}=5$--32 (for d.o.f. changing from 1019 to 1007, correspondingly), 
with the high-$N_{\rm H}$
low-$kT$ solution always with the lower ${\rm CSTAT}$. We interpret it as an indication that
 these two parameters
are indeed degenerate in our modelling, with a slight hint towards the high-$N_{\rm H}$ low-$kT$ solution.

\begin{figure}
\includegraphics[width=85mm]{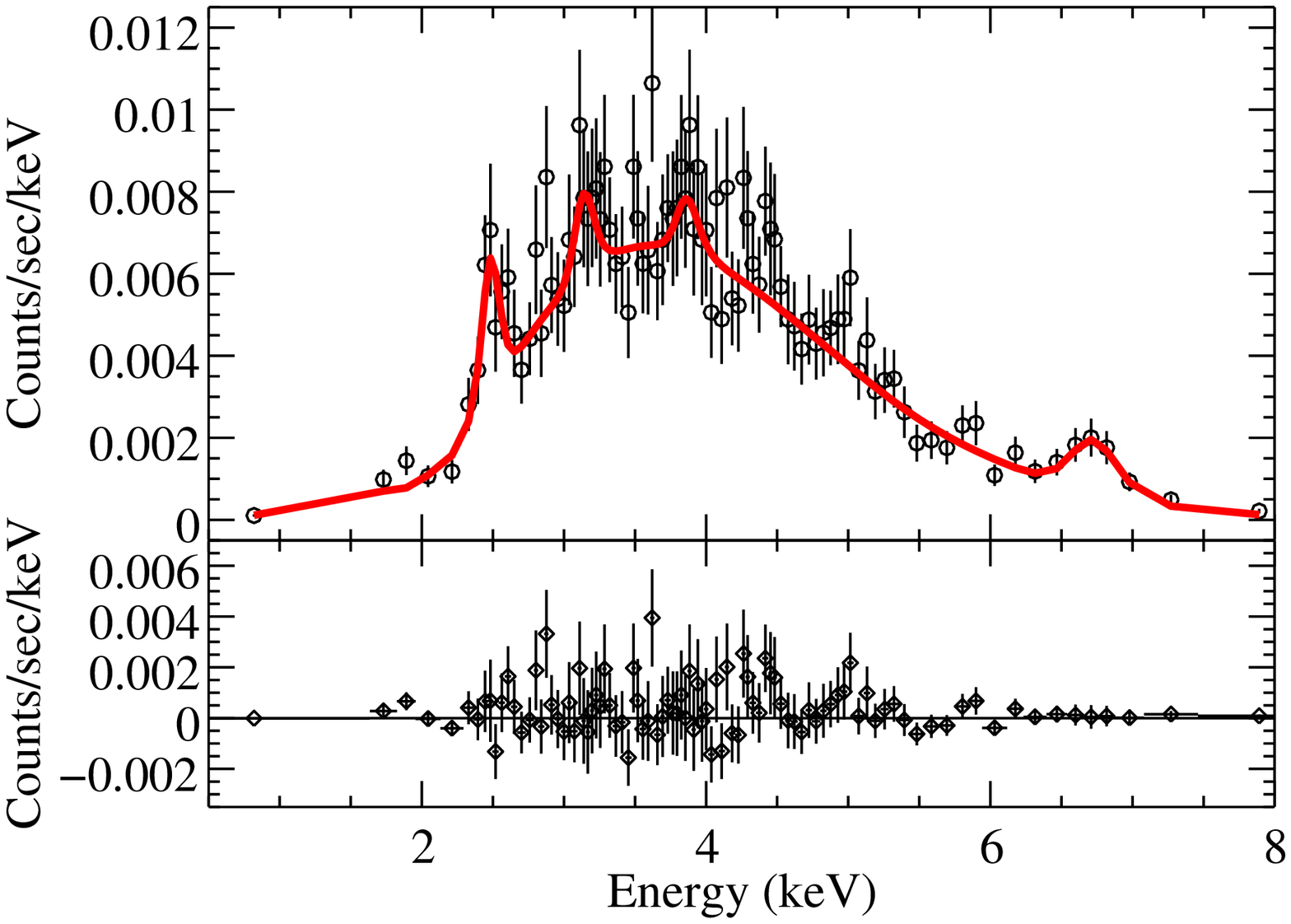}
\includegraphics[width=85mm]{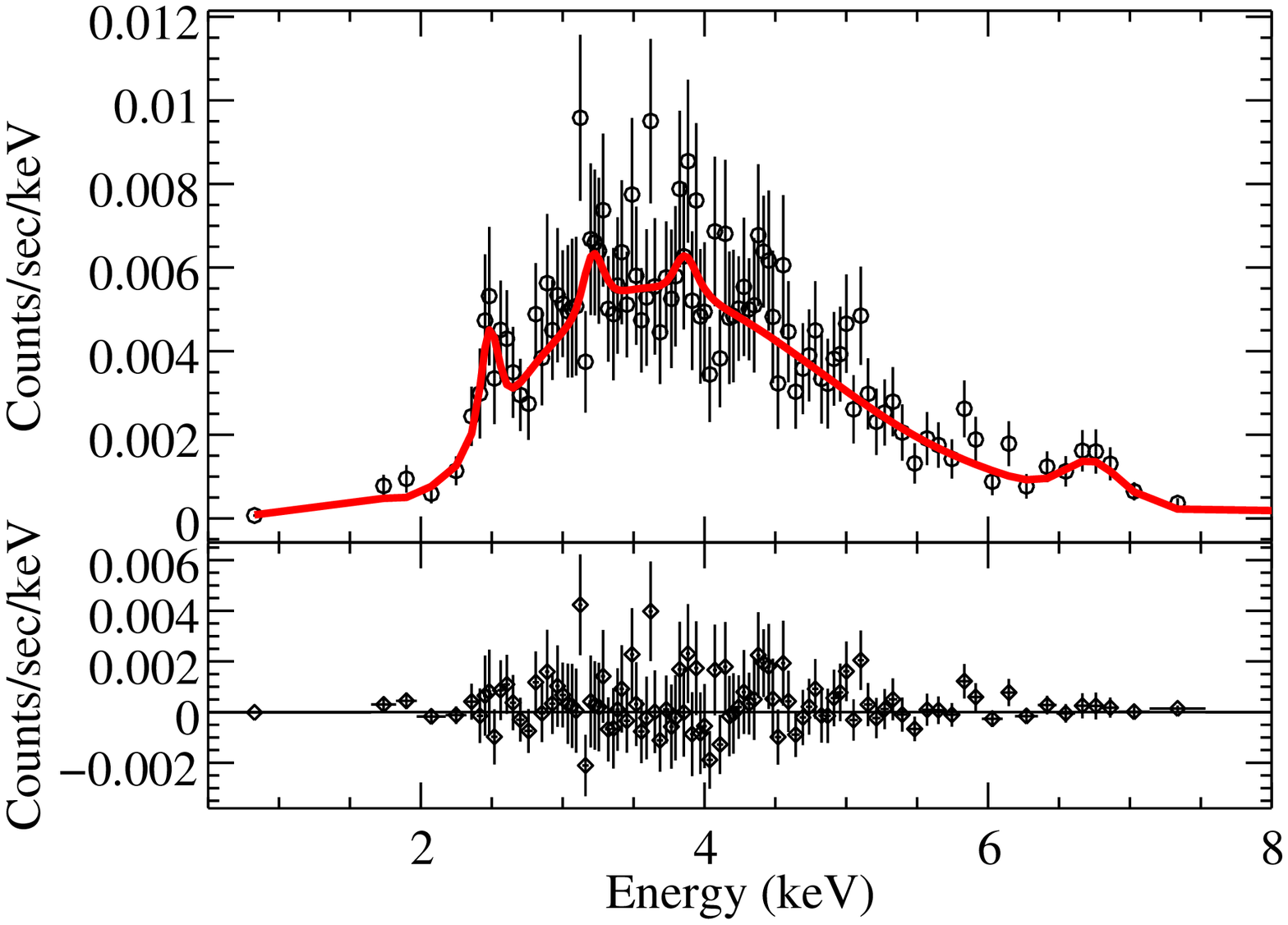}
\caption{Results of spectral fitting of $5''$ region - upper panel for distant
background, lower panel for annulus $6''-18''$. The total model 
marked with  a solid red line is a sum of Galactic absorption (tbabs),
thermal bremsstrahlung and four Gaussian lines for Fe, S, Ar and Ca emission.} 
\label{fig:chi} 
\end{figure}

Tab.~\ref{tab:simpsd} shows also the
the best-fit model parameters for two additional fits with 
total model {\sc tbabs + bremss} + 4 lines, using  the default Sherpa statistic (chi2gehrels), 
and background subtracted data grouped requiring that $S/N=3$. We performed these fits in order to
check how different ways of treating the background affect our analysis. The third and fourth rows of
the table presents results obtained using the local and distant backgrounds, respectively.
The corresponding plots are presented in Fig.~\ref{fig:chi}. 

The change of the statistics does not change the results if the same 
(distant) background is used. All model
parameters are consistent within their uncertainties. In the case with the local background, we observe
a further decrease in the plasma temperature to a value $kT = 2.24^{+0.07}_{-0.07}$\,keV and increase
in the absorbing column, $N_{\rm H} = 10.52^{+0.42}_{-0.39} \times 10^{22}$\,cm$^{-2}$; these values are not consistent
at $2\sigma$ level with the estimates from the other models including all 4 emission lines.

We conclude that the continuum is well fitted by a thermal bremsstrahlung model indicating 
temperatures for the plasma around Sgr~A* in the 2.2--2.7\,keV range, depending on the choice of background.
These values are lower than that obtained by \citet{wang2013} with the same model for the $1.5''$ Sgr A*
spectrum; they quote $3.5(3.0, 4.0)$\,keV. 
The EW of the iron line we find is in the $0.9-1.2$\,keV range, but its uncertainties are high,
due to the uncertainty in deriving the relative contribution of the continuum and line, the number 
of model parameters, and the quality of the data. Within the uncertainties, our EW is
consistent with that estimated by \citet{wang2013} at the $1\sigma$ confidence level 
(EW$=691 (584, 846)$\, eVs).
Our data are of not sufficient quality to study the Fe line properties 
(e.g. its evolution with the distance from Sgr A*)
in details and for this purpose we address the reader to the work of \citet{wang2013}.
The absorbing column in \citet{wang2013}, $10.1(9.4, 11.1)\times 10^{22}$\,cm$^{-2}$,
is also consistent with our results at the $1\sigma$ confidence level. 
Note, however, that this value of $N_{\rm H}$ is
inconsistent with the absorbing column they find for the flaring Sgr A* spectrum, 
and based on this they tend to discard
this phenomenological fitting as unphysical and continue with the development of their RIAF model, 
where they find plasma temperature $\sim$1\,keV. Our value of $kT$ is somewhat intermediate 
between those of \citet{wang2013} resulting form their phenomenological and RIAF models.
However, in the following, we show alternative to the \citet{wang2013} fitting 
of the surface brightness profile. Instead of computing emissivity from the gas 
with temperature and density profile analytical expressions,  which mimic RIAF dynamical model as done by \citet{wang2013}, we fit dynamical 
Bondi flow directly to the luminosity map, without any profile functions.

\section{Hot flow around Sgr~A*}
\label{sec:flow}

{\it Chandra} ACIS-I camera spatially resolves 
close vicinity of the central supermassive black hole  and the 
observations constrain the model of hot plasma around Sgr~A*. 
Previously, \citet{shcherbakov2010} analyzed  953 ksec {\it Chandra}
data and they have  fitted model of hot plasma 
interacting with stars up to $5''$ from the Sgr~A* complex. 
They have provided density and temperature radial profiles of the hot plasma at the outer 
radius of the flow
to be equal: $T = 2$ keV and $n_{\rm e}= 60 $ cm$^{-3}$. 
More recently, \citet{wang2013} reported 3 Msec observations of Sgr~A*, 
where they have concluded that  RIAF model, when the spectrum is 
fitted by calculating emissivity from RIAF temperature and density profiles,  
provides an excellent spectral fit indicating 
the plasma temperature to be 1 keV.
Unfortunately, the authors did not fit surface brightness 
profile to the data.
 
In the data extracted by us, the collected counts per pixel square are above 30 in the 
center and a few at $10''$ (Fig.~\ref{fig:profil} red line). 
Below we present the fit to the surface brightness profile
with the model of the hot Bondi flow. The purpose of our studies is to 
show to which extend the Bondi flow works. In principle in the very centre 
star influence can be inefficient as discussed by \citet{shcherbakov2010}.
In any of such models or RIAF model with outflow, there exists a stagnation radius, which 
indicates the border between the matter inflowing or outflowing from the very center. 
At the stagnation radius the Bondi flow stops to be valid, as we show in the section below.

\subsection{Dynamical model of the hot accretion flow}
\label{sec:bondi}

We compute model brightness profile based on a standard spherical accretion
flow theory \citep{bondi1952}. 
The plasma density, temperature, and velocity radial profiles are
  obtained by solving the general Bondi equations 
  \citep[the conservation equations given in][equations G.21 and G.22]{shapiro83}. 
   In particular we solve the following Bernoulli equation:
\begin{equation}
h \sqrt{1-\frac{2}{r}+u^{\rm r}} = h_{\rm s} \sqrt{1-3*cs_{\rm s}^2}
\end{equation}
where fluid enthalpy is $h=1/(1-cs^2/(\gamma_{\rm ad}-1))$, $cs$ is the speed of
sound, $u^{\rm r}$ is the fluid velocity, and the right-hand-side is the
integration constant, found by taking the $h$ and $cs$ at the sonic
radius, denoted with underscore {\it s}. The speed of sound at the sonic radius $r_s$ is given by
$cs_{\rm s}^2=1/(2*r_{\rm s}-3)$  \citep[see Eq.G17 in][]{shapiro83}. 
The adiabatic index is assumed to be $\gamma_{\rm ad}\approx 5/3$.  Although we use equations that
generalize Bondi equations for the Schwarzschild spacetime, our solutions at
large distances from the black hole naturally converge to the classical Bondi
solution.

The above Bondi model has two (except $\gamma_{\rm ad}$ which is fixed in our
  model) free parameters: density $n_{\rm e}^{\rm out}$ and temperature
of plasma $T_{\rm e}^{\rm out}$, both at the outer radius $R_{\rm out} \geq 10''$. These 
variables are assumed far away from the black hole, and so they are equivalent
to the ``infinity'' values (i.e. $n_{\rm e}^{\rm out} \equiv n_{\infty}$, 
$T_{\rm e}^{\rm out} \equiv T_{\infty}$).
The mass accretion rate onto the black hole in the Bondi models in the units of g s$^{-1}$ is:
\begin{equation}
\dot{M}=4\pi\lambda_s \frac{(G M_{\rm BH})^2 \rho_{\infty}}{c_{\rm s,\infty}^3}
\label{eq:mdot}
\end{equation}
where $\lambda_{\rm s}$ is a function of adiabatic index $\gamma_{\rm ad}$ (for our 
$\gamma_{\rm ad}\approx 5/3$, $\lambda_{\rm s}=0.25$). We integrate the model
equations from $R_{\rm out}$ down to the black hole horizon. The list of
models explored in this work is given in the Tab.~\ref{tab:BSmodels}.

\begin{table*}
\caption{List of spherical accretion model parameters ($n_{e,\infty}$ and
  $T_\infty$) with the corresponding mass accretion rate  (Eq.\ref{eq:mdot}), 
 the speed of sound at the infinity, and Faraday Rotation measures for assumed
radial, coherent, moderate ($\beta = 1$)  and weak ($\beta = 10^7$)  magnetic fields. 
Fifth and sixth columns show RM$_{\rm GC}$ computed towards Galactic Center, 
while seventh and eight columns represent  RM$_{\rm Pulsar}$ computed
towards pulsar PSR J1745-2900, located $3''$ away from Sgr~A* in the projected distance.
For comparison, the measured value of RM$^*_{\rm GC} = 5.6 \pm 0.7 \times 10^5$ rad/m$^2$ 
\citep{marrone2007}, and 
RM$^*_{\rm Pulsar}=6.696 \pm 0.004 \times 10^4$ rad/m$^2$  \citep{eatough2013}.}
\label{tab:BSmodels}
\begin{center}
\begin{tabular}{cccccccc}
\hline
 $\rho_{\infty}$ & $T_{\infty}$ & $\dot{M}$ & 
$c_{s,\infty} $ & RM$_{\rm GC}$ ($\beta=1$) & RM$_{\rm GC}$ ($\beta=10^7$) & 
  RM$_{\rm Pulsar}$ ($\beta=1$) &  RM$_{\rm Pulsar}$ ($\beta=100$) \\
  ${\rm [cm^{-3}]}$ &  [keV] & $ {\rm [M_\odot yr^{-1}]}$ & ${\rm [cm s^{-1}]}$ & 
  ${\rm [rad/m^2]}$ &  ${\rm [rad/m^2]}$ & ${\rm [rad/m^2]}$
   & ${\rm [rad/m^2]}$  \\

\hline
1 & 1 & $4.6 \times 10^{-7}$ & $3.9 \times 10^7$ & $4.4 \times 10^8$ & $ 1.4 \times 10^5$ & $4.5 \times 10^2$ &  40 \\
1 &3.5& $7.1 \times 10^{-8}$ & $7.4 \times 10^7$ & $2.6 \times 10^7$ & $ 8.4 \times 10^3$ & $6.6 \times 10^2$ &  70 \\
1 & 4 & $5.8 \times 10^{-8}$ & $7.9 \times 10^7$ & $2.0 \times 10^7$ & $ 6.2 \times 10^3$ & $7.1 \times 10^2$ &  70 \\
1 & 5 & $4.1 \times 10^{-8}$ & $8.9 \times 10^7$ & $1.2 \times 10^7$ & $ 3.8 \times 10^3$ & $7.8 \times 10^2$ &  80 \\
1 & 6 & $3.1 \times 10^{-8}$ & $9.7 \times 10^7$ & $7.9 \times 10^6$ & $ 2.5 \times 10^3$ & $8.4 \times 10^2$ &  80 \\
1 & 8 & $2.0 \times 10^{-8}$ & $1.1 \times 10^8$ & $4.1 \times 10^6$ & $ 1.3 \times 10^3$ &  $9.6 \times 10^2$ & $1.0 \times 10^2$ \\
1 & 16& $7.2 \times 10^{-9}$ & $1.5 \times 10^8$ & $9.0 \times 10^5$ & $ 3.0 \times 10^2$ & $1.3 \times 10^3$ & $1.3 \times 10^2$  \\
\hline
18.3 & 3.5 & $1.3 \times 10^{-6}$ & $7.4 \times 10^7$& $2.1 \times 10^9$ & $6.6 \times 10^5$& $5.2 \times 10^4$ & $ 5.2 \times 10^3$ \\
18.4 & 3.5 & $1.3 \times 10^{-6}$ & $7.4 \times 10^7$& $2.1 \times 10^9$ & $6.6 \times 10^5$ & $5.3 \times 10^4$ & $ 5.3 \times 10^3$\\
\hline
35 & 1 & $1.6 \times 10^{-5}$ & $3.9 \times 10^7$& $9.1 \times 10^{10}$ &$2.9 \times 10^7$& $ 9.3 \times 10^4$ & $ 9.3 \times 10^3$ \\
35 &3.5& $2.4 \times 10^{-6}$ & $7.4 \times 10^7$& $5.5 \times 10^9$ & $1.7 \times 10^6$ & $ 1.4 \times 10^5$ & $ 1.4 \times 10^4$ \\
35 & 4 & $2.0 \times 10^{-6}$ & $7.9 \times 10^7$& $4.1 \times 10^9$ & $1.3 \times 10^6$ & $ 1.5  \times 10^5$ & $ 1.5 \times 10^4$ \\
35 & 5 & $1.4 \times 10^{-6}$ & $8.9 \times 10^7$& $2.5 \times 10^9$ & $7.8 \times 10^5$ & $ 1.6 \times 10^5$ & $ 1.6 \times 10^4$ \\
35 & 6 & $1.1 \times 10^{-6}$ & $9.7 \times 10^7$& $1.6 \times 10^9$ & $5.2 \times 10^5$ & $ 1.7 \times 10^5$ & $ 1.7 \times 10^4$\\
35 & 8 & $7.2 \times 10^{-7}$ & $1.1 \times 10^8$& $8.6 \times 10^8$ & $2.7 \times 10^5$ & $ 2.0 \times 10^5$ & $ 2.0 \times 10^4$ \\
35 & 16& $2.5 \times 10^{-7}$ & $1.5 \times 10^8$& $1.8 \times 10^8$ & $5.8 \times 10^4$ & $ 2.8 \times 10^5$ & $ 2.8 \times 10^4$\\
\hline
60 & 1 & $2.8 \times 10^{-5}$ &$3.9 \times 10^7$& $2.0 \times 10^{11}$ & $ 6.4 \times 10^7$ &  $ 2.1 \times 10^5$ & $ 2.1 \times 10^4$\\
60 &3.5& $4.2 \times 10^{-6}$ &$7.4 \times 10^7$& $1.2 \times 10^{10}$ & $ 3.9 \times 10^6$ &  $ 3.1 \times 10^5$ & $ 3.1 \times 10^4$\\
60 & 4 & $3.4 \times 10^{-6}$ &$7.9 \times 10^7$& $9.1 \times 10^9$ & $ 2.9 \times 10^6$ & $ 3.3 \times 10^5$ & $ 3.3 \times 10^4$\\
60 & 5 & $2.5 \times 10^{-6}$ &$8.9 \times 10^7$& $5.5 \times 10^9$ & $1.7 \times 10^6$ & $ 3.6 \times 10^5$ & $ 3.6 \times 10^4$\\
60 & 6 & $1.9 \times 10^{-6}$ &$9.7 \times 10^7$& $3.7 \times 10^9$ & $1.2 \times 10^6$ & $ 3.9 \times 10^5$ & $ 3.9 \times 10^4$\\
60 & 8 & $1.2 \times 10^{-6}$ &$1.1 \times 10^8$& $1.9 \times 10^9$ & $ 6.1\times 10^5$ & $ 4.4 \times 10^5$ & $ 4.5 \times 10^4$\\
60 & 16& $4.6 \times 10^{-7}$ &$1.5 \times 10^8$& $4.1 \times 10^8$ & $1.3\times 10^5$ & $ 6.2 \times 10^5$ & $ 6.2 \times 10^4$ \\
\hline
\end{tabular}
\end{center}
\end{table*}

To create synthetic brightness profiles, we first generate images of the inner
region of the GC $20'' \times 20''$ using a ray tracing method. 
Our scheme performs radiation
transfer through the three dimensional spherically symmetric 
accretion flow model and records the
radiation flux density at each of the detector pixels. In our modelling, we assume 
that the X-ray radiation is produced by bremsstrahlung.  We solve the 
radiation transfer equation along many rays, using
the following bremsstrahlung emissivity function: 
\begin{equation}
\epsilon_{\rm ff,\nu}=6.8 \times 10^{-38} \,
n_{\rm e}^2 \,\, T_{\rm e}^{-1/2} \, \exp \left( \frac{-h\nu}{kT} \right).
\end{equation}
The plasma absorptivity is also taken into account and it is derived
from the Kirchhoff's law of thermal radiation: 
$\alpha_{\rm ff,\nu}=\epsilon_{\rm ff,\nu}/B_\nu$, where $B_\nu$ is the Planck
function.

To convert the theoretical flux to counts detected by ACIS-I {\it Chandra} 
CCD, the model fluxes are convolved with the instrument response. 
The total number of counts detected during the exposure time $T_{\rm exp}$ is:
\begin{equation}
C_{\rm tot} = \int T_{\rm exp} \, F_{\rm \nu} \,
A_{\rm eff}(\nu) \, \nu^{-1} \, exp(-abs(\nu) \, N_{\rm H}) \, d\nu ,
\end{equation}
where $\nu$ is a frequency corresponding to energies from 0.5 up to 8 keV.
Additionally, we assume that absorption also takes place in the ISM between
the outer boundary of accretion model and the detector. Hence, we multiply the
flux by additional exponential function, and $N_{\rm H}=10^{23}$ cm$^{-2}$ is 
a typical column density of the ISM. Effective area of the instrument 
$A_{\rm eff}(\nu)$ and absorption
coefficient in the ISM  abs$(\nu)$ are energy dependent; we adopt the same
dependencies as in \citet[][priv. communication]{shcherbakov2010}. 

The final modelled emissivity profiles are obtained by mapping the
square images ($20'' \times 20''$) onto 40 concentric rings  (the same 
rings that were used for extracting the data.  
The total count rate collected on each ring is 
divided by that ring area (ring area is given in a units of pixels). 

Thus computed profiles have to be convolved with the PSF derived for our 
particular observation in Sec.~\ref{sec:pro}. 
Examples of convolved models of Bondi hot accretion  are shown
in Fig.~\ref{fig:fit}  by dotted and short dashed lines for 
two outer temperatures equal 1 and 3.5 keV respectively. 
For fitting procedure, we have computed a grid of models with different 
temperatures and densities ranging in $T_{\rm e}^{\rm out}=1-16$ keV and 
$ n_{\rm e}^{\rm out} = 1-60$ cm$^{-3}$  (see Tab.~\ref{tab:BSmodels}).

\subsection{The best model of the surface brightness profile} 
\label{sec:wang}

We fit the model emissivity profile from the Bondi accretion flow 
convolved with instrument response and with the PSF to the observed 
surface brightness profile. 
In Fig.~\ref{fig:bright} (lower panel) the residuals of fitting up to $10''$ 
are shown. The fit is not statistically acceptable, showing that the
uniform Bondi flow is not a good representation for the hot plasma  at distances 
up to $10''$ from Sgr~A*.
In previous analysis by \citet{shcherbakov2010} a dynamical model 
of the hot plasma interacting with stars was fitted to the surface brightness 
profile up to the $5''$ from GC.  
 Additionally, \citet{wang2013} pointed out that Bondi accretion 
flow ends at the radius $\sim 3.6 \times 10^5 R_{\rm Schw}$ which is 
 $3.8''$ in the case of $M=4.4 \times 10^6 M_\odot $ black hole.

\begin{figure}
\hspace{0.2cm}
  \includegraphics[angle=0,width=8.7cm]{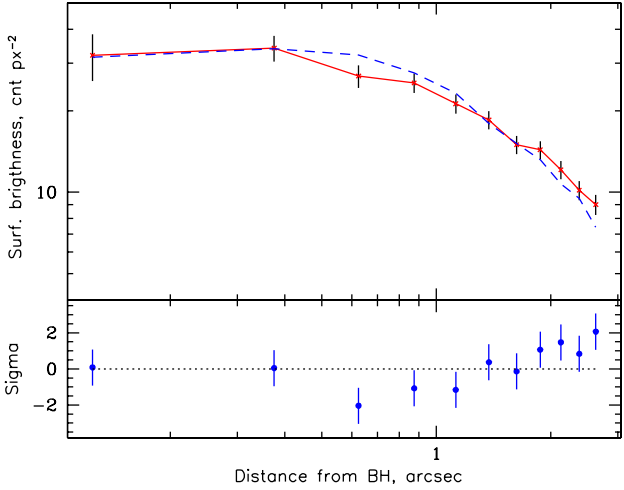}
\caption{Surface brightness profile up to $3''$ away from Sgr~A*. 
Red points represent observed profile, and are connected by continuous line. 
The best fit model with  $T_{\rm e}^{\rm out}=3.5 \pm 0.3$ keV and 
$n_{\rm e}^{\rm out}=18.3 \pm 0.1$ cm$^{-3}$
is presented by blue short dashed line, and the fit residuals are presented
in the lower panel.}
\label{fig:fit}
\end{figure}

\begin{figure}
\includegraphics[angle=0,width=8.5cm]{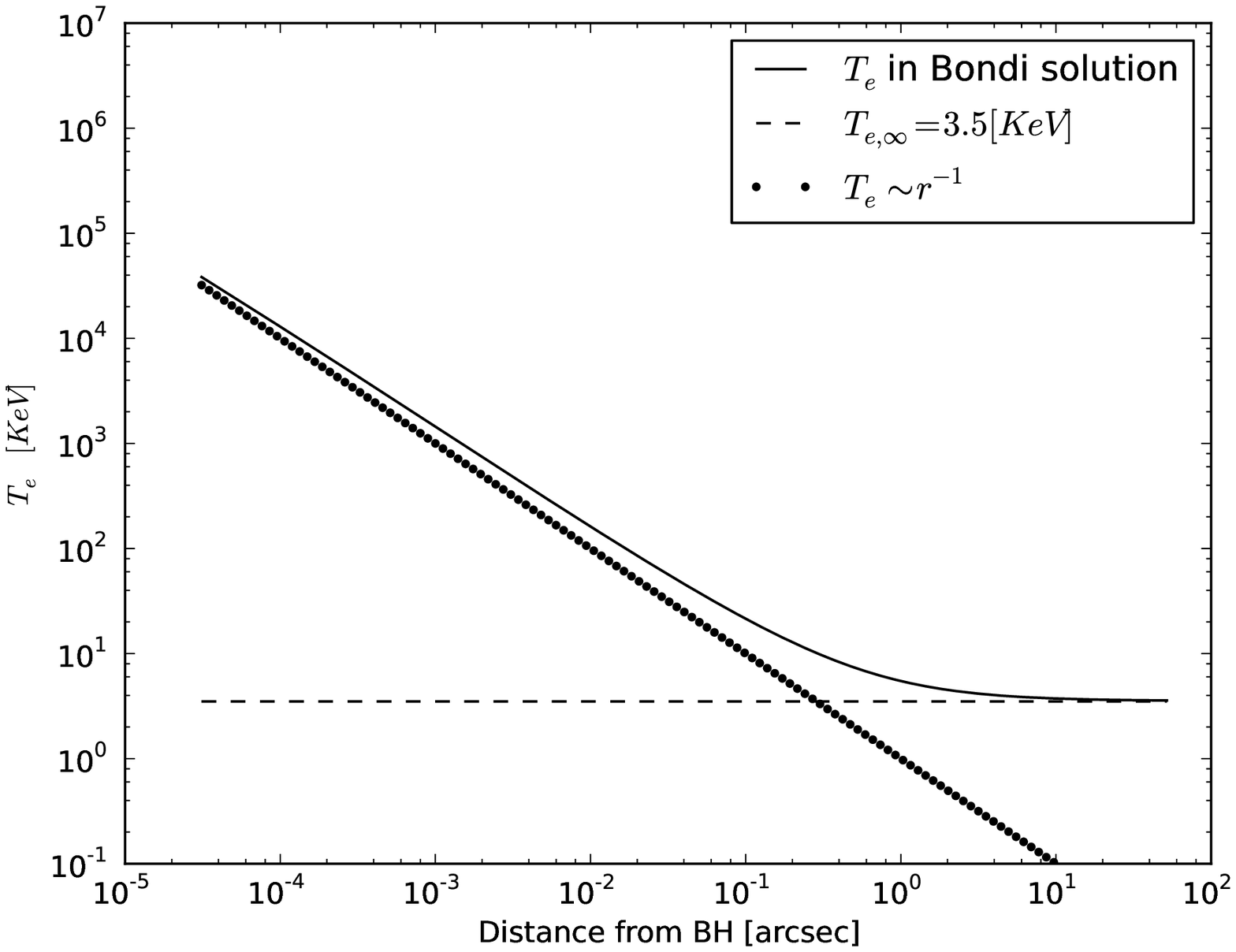}
\includegraphics[angle=0,width=8.5cm]{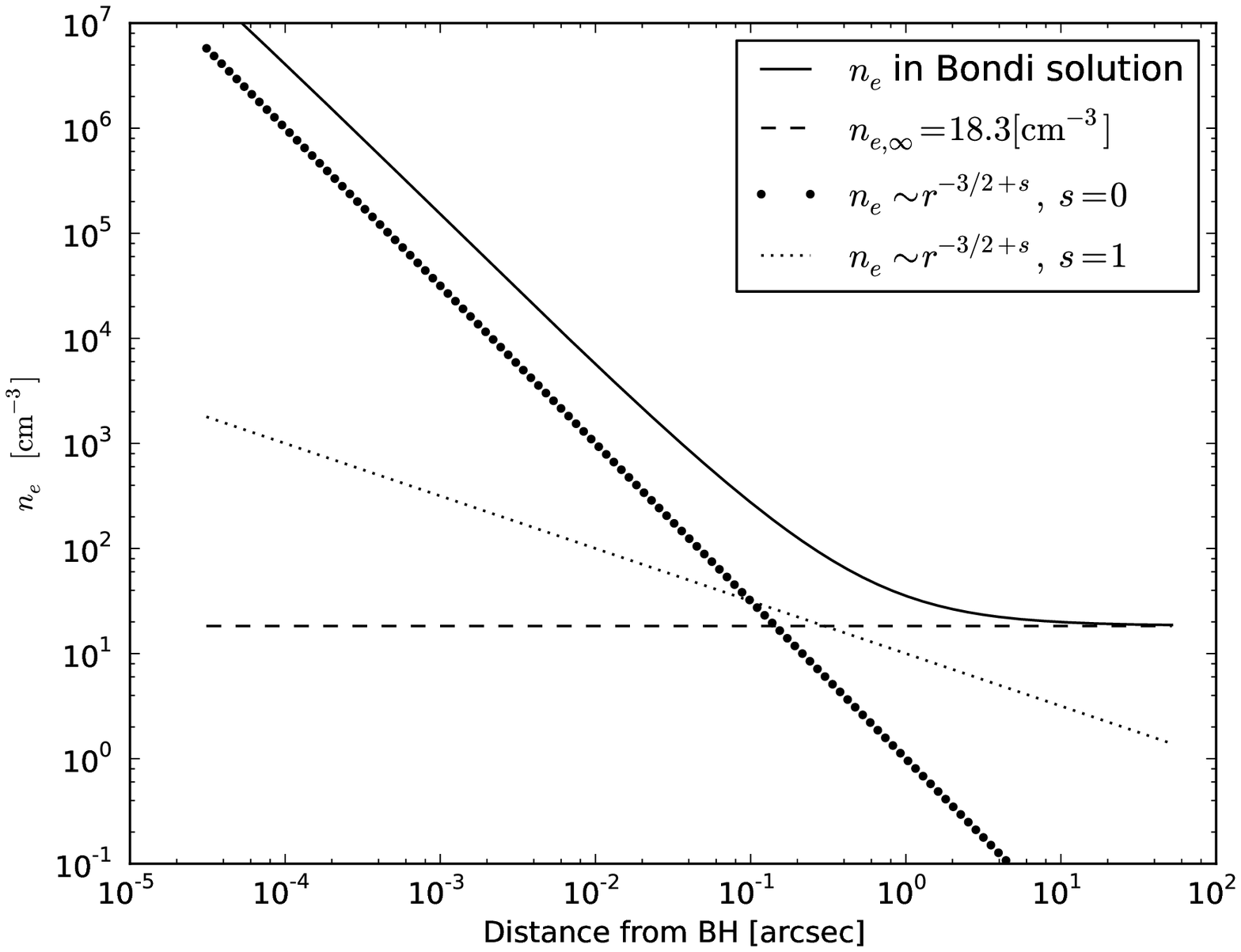}
\caption{Temperature and density profiles around Sgr~A* for our best fitted
model $T_{\rm e}^{\rm out}=3.5 \pm 0.3$ keV and $n_{\rm e}^{\rm out}=18.3 \pm 0.1 $ cm$^{-3}$, 
upper and lower panel 
respectively. The asymptotic values of $T_{\rm e}^{\rm out}$
and $n_{\rm e}^{\rm out}$ are marked as a dashed lines. 
Additionally, we show the radial dependencies of the temperature and density 
in RIAF model without (s=0) and with (s=1) an outflow. The normalization of
RIAF models is arbitrary. }
\label{fig:temp}
\end{figure}

For this reason, we restricted our modelling to smaller radii and 
obtained a good fit up to $3''$ from  Sgr~A*. The best theoretical fit profile 
is shown with a dashed line in the upper panel of Fig.~\ref{fig:fit}, and the parameters of 
the spherical accretion model are: $n_{\rm e}^{\rm out}=18.3 \pm 0.1$ cm$^{-3}$ 
and $T_{\rm e}^{\rm out}=3.5 \pm 0.3 $ keV.  The corresponding integrated 
mass accretion rate is $\dot{M}=1.3 \times 10^{-6}$\, M$_\odot$~yr$^{-1}$ . 
The reduced $ \chi^2$ of  our best fit is
1.19. Residuals are presented in the lower panel of Fig.~\ref{fig:fit}. 
All data points are within 2 sigma from the model. 

From the best fitting model we can estimate profiles of 
temperature and density of the hot plasma. 
We present them in Fig.~\ref{fig:temp}  with solid lines. As a result of our fitting 
we found the values of density and temperature at the distance 
of $1.5''$ to be $ n_{\rm e} = 29.2$ cm$^{-3}$ and 
$T_{\rm e} = 4.8$ keV, respectively. These values are different than 
those obtained by \citet{shcherbakov2010} ($n_{\rm e} = 130$ cm$^{-3}$ and 
$T_{\rm e} = 2$ keV).

The mass accretion rate from our Bondi best fit model is 10-1000 times larger 
in comparison to the mass accretion rate limits set by the measurements of the 
Faraday rotation near the central black hole 
$10^{-9}<\dot{M}<10^{-7}$ ~M$_\odot$~yr$^{-1}$  \citep{bower2005,marrone2006,marrone2007}. 
Nevertheless, the derived mass accretion rate 
is always model dependent and the direct comparison is meaningless. 
The above limits computed by \citet[][see Eq. 8 in this paper]{marrone2006} 
strongly depend how the Faraday rotation measure (RM) relates to the accretion flow and 
magnetic field projected on the line of integration. 
To avoid any inaccuracies, we 
present here RM computed from our models according to the formula:
RM=$8.1 \times 10^5 \int n_e {\rm B} dl$, where electron density $n_e$ is in cm$^{-3}$, 
the path length $dl$ is in pc, and the magnetic field ${\rm B}$ - in Gauss
\citep{gardner66}, the last defined by standard formula $ {\rm B}^2= 8 \pi P_{\rm
  gas}/ \beta$, where $\beta$ is the plasma parameter.

There is no direct constraint on the magnetic field value in Sgr~A*, therefore we consider the 
case of moderate, and very weak magnetic field. We present 
RM  computed towards the  GC from all our models in 
Tab.~\ref{tab:BSmodels}. Fifth column is for  $\beta = 1$ (magnetic field in
equipartition  with the gas), while sixth column is for 
$\beta = 10^7$. For $\beta = 1$, RM$_{\rm GC}$  for the best fitted model 
does not agree with measured value 
RM$^*_{\rm GC} = 5.6 \pm 0.7 \times 10^5$ rad/m$^2$ \citep{marrone2007}.
But it fully agrees, if we assume that the magnetic field in the radial direction
is very weak i.e. $\beta = 10^7$. 
We therefore admit, that our best fit model is not well constrained within approximately
$100  R_{\rm Schw}$ or the magnetic field in radial direction is very weak.
When we start the integration of our model from $5 \times 10^{-3}$$''$ (which corresponds to the
distance of $464 \, R_{\rm schw}$), 
the resulting RM$_{\rm GC} = 6.91 \times 10^5$ rad/m$^{2}$ with magnetic field being in equipartition. 
In this case, our result is in quit good agreement with observations, especially that in the paper
by \citet{marrone2006} (Fig. 4) the integration of models was typically done from $300 \, R_{\rm schw}$, and than 
comparred to the data.

For parameter  $\beta = 10^7$, the magnetic field in the center of the flow is
${\rm B} = 5 \times 10^{-2}$ Gauss, which is not in agreement with recent found that 
the strength of ${\rm B}$ in GC is around hundred Gauss \citep{eatough2013}. Nevertheless, this value was 
given by authors after they have measured RM toward the pulsar PSR J1745-2900 near the GC. 
Since all model computations are not precise very close to the black hole, we decided 
to compare the modelled RM towards the above pulsar, located $3''$ in the 
projected distance from the black hole. Such value was reported recently to be equal
RM$^*_{\rm Pulsar} =6.696 \pm 0.004 \times 10^4$ rad/m$^2$  \citep{eatough2013}.
We project our density profile and magnetic field on the direction towards
the pulsar  and compute modelled RM$_{\rm Pulsar}$ in case of $\beta = 1$ and $\beta = 100$. 
We present results in Tab.~\ref{tab:BSmodels}, seventh and eight column respectively.
 
The RM$_{\rm Pulsar}$ for the best fitted model agrees very well 
with the observed value for $\beta = 1$.  For such value of $\beta$ parameter
the magnetic field towards Sgr~A* is ${\rm B} = 157$ Gauss, from our best fitted model. 
We conclude, that our model density and temperature profiles, and hence mass accretion 
rate of about $10^{-6}$  M$_\odot$~yr$^{-1}$ around $3''$ are 
consistent with RM measured toward the GC pulsar. It is possible that the
mass accretion rate is reduced at distances closer to the black hole. However,
these outflows cannot be accounted for in our simple spherically symmetric model
that is fitted to spherically averaged brightness profile.

For comparison Fig.~\ref{fig:temp} also shows the radial profiles of
temperature and gas number density from analytical radiativelly inefficient
accretion flow 
(RIAF) models \citep{shapiro76,rees82}. In such models 
the density profile is given by $n \sim r^{(-3/2+s)}$ where $s$ is parameter
that if positive then the RIAF launches  an outflow. In Fig.~\ref{fig:fit} 
lower panel, we plot density profile for
$s=0$ (no outflow) and $s=1$, by big and small dots respectively. 
The density normalization is arbitrary. In RIAF the
temperature profile follows $T \sim r^{-\theta}$ where parameter 
$0<\theta<1$ (the same figure upper panel).
Typically in RIAF $\theta=1$ , i.e. temperature has a virial profile \citep{wang2013}. 
In Bondi spherical accretion below the sonic point the density
profile follows the same radial dependency as in RIAF without the outflow
(i.e. s=0), and the temperature is virial. However, the Bondi 
solutions at large radii naturally connects to constant values, and both,
density and temperature profiles deviate from the $n \sim r^{-3/2}$ and
$T \sim r^{-1}$ laws, despite of the fact that there is no outflow in this
solution.

 Our result does not contradict other accretion flow models considered in case 
of Sgr~A*. It shows the position of stagnation radius to be at $3''$.
Outside this distance, profiles derived by us may not work.

\section{Conclusions}
\label{sec:summary}

In this paper we presented 134 ks {\it Chandra} ACIS-I observations of the GC 
containing the Sgr~A* region. After analysing 
flux images in the continuum (0.5 -- 8 keV) and iron line (6.3 -- 7.0 keV) bandpasses, 
we have extracted spectrum  of  $5''$ circular region around Sgr~A*. 
Our spectral model consists of a thermal plasma for the continuum plus Gaussian 
profiles for the K$_{\alpha}$ emission of Fe, S, Ar and Ca lines. 
The continuum is well fitted by a thermal bremsstrahlung emission indicating temperatures
for the plasma around Sgr~A* in the 2.2 -- 2.7 keV range, depending on the choice of 
background. The corresponding absorbing column densities towards the source are 
10.5 -- 9.2$\times 10^{22}$ cm$^{-2}$, and the corresponding 
EWs of Fe$_{K_{\alpha}}$ line are 1.19 -- 0.91 keV. 
Our spectral fitting indicates lower temperature by about 0.5 keV than this 
obtained by   \citet{wang2013} with the same model. 
Nevertheless, the strength of iron line consistent with that reported by  \citet{wang2013} 
at the $1 \sigma$ confidence level. 

Furthermore, we studied the brightness profile of the hot plasma 
in the vicinity of the Sgr~A* black hole.  We found that the classical 
Bondi accretion can reproduce observed brightness profile up to $3''$. 
This result implies strong constrains on the position of stagnation point always 
considered to be present in the complicated flow around Sgr~A*.
The best fit model for surface brightness profile requires that the hot plasma 
outside the flow to have temperature $T_{\rm e} = 3.5 \pm 0.3$ keV and 
density $n_{\rm e} =18.3 \pm 0.1$ cm$^{-3}$.

Sgr~A* has been recently resolved down to $1.5''$ and studied by 
\citet{wang2013}. They have reported an X-ray image of the source based on a  
3 Msec {\it Chandra} observation. The best fitted model was RIAF 
(radiatively inefficient accretion flow), but 
they did not show any resulting temperature and density profiles.   
Only a general description for the temperature dependence on the distance 
was given as: $T \propto  r^{- \theta}$ where $ \theta \gtrsim 0.6 $. 
Additionally, they concluded from the weak line of H-like iron
Fe XXVI, that the amount of gas at temperatures $\sim 9 $ keV is very small. 
Our temperature profile predicts 
the temperature of 9 keV at the distance $0.4''$ from the 
black hole. Such plasma would have number density equal to 70 cm$^{-3}$.
Both temperature profiles, RIAF and Bondi, agree up to $0.1''$, as it is 
seen in Fig.~\ref{fig:temp}. But Bondi flow fitted by us predicts higher 
density by one order of magnitude. 

All models show different temperatures at the outer parts of the flow
$T_{\rm e}$ = 1, 2, and 3.5 keV  respectively in the RIAF model \citep{wang2013}, 
hot plasma fed by stars  \citep{shcherbakov2010}, and the only Bondi 
flow reported in this paper.  
Additionally, spectral fitting done by us and \citet{wang2013} indicates slightly different 
temperatures (2.7 and 3.5 respectively). 
Those results show a big complexity of matter structure around Sgr~A*, 
and additional indications should be taken into account when studying hot plasma 
in this region. For instance  \citet{wang2013} also found the He-like Fe K$_{\alpha}$ line.
Therefore, we argue, that up to $3''$ Bondi flow works, but further away hot matter 
may be in the form of two-phase medium. 
We speculate that the Fe K$_{\alpha}$ line emission at $6.69$ keV can be attributed 
to the reflection from clouds of the warm gas. This illuminated radiation originates from the 
very innermost nucleus and depending on the luminosity state of Sgr~A*, 
can provide strong clumpiness and trigger formation of such hot clouds, as proposed 
by \citet{rozanska14}.
However, it is critically important to confirm that the Bondi model fitted here does not 
overpredict the strength of the FeXXVI line. \citet{wang2013} concluded that due to the 
weakness of FeXXVI line, there must be a little gas at temperatures larger than 9 keV. 
In our Bondi model the gas temperature reaches 9 keV at $0.4"$ (compared to $~0.1"$ in the RIAF 
model). If the strength of this line computed from our best fitted model of Bondi flow
disagrees with the observed one, this may rule out of our model despite its agreement 
with the surface brightness profile.
To fully check our predictions we plan to construct the model of thermal plasma. 
Also, better spectrum will be needed for this kind of analysis.

The mass accretion rate from our Bondi best fit model is $1.3 \times 10^{-6}$~
M$_\odot$~yr$^{-1}$. This is out of 
the range of accretion rates predicted from 
measurements of the Faraday rotation (RM) \citep{marrone2007}.
Nevertheless, we point out that the derived mass accretion rate limits 
are always model dependent. Especially we can modify them by changing the 
structure and strength of a magnetic field. 
For our best fitted model the computed RM$_{\rm GC}$ agrees with observed one 
toward Galactic
Center with assumption of very weak radial magnetic field. 
Moreover, RM$_{\rm Pulsar}$ computed from the best fit model
towards the pulsar PSR J1745-2900 located at $3''$ projected distance 
from GC is in very good agreement with observations assuming moderate magnetic
field, $\beta=1$. This result clearly shows that any limitation on accretion rate from Faraday rotation
strongly depends on the model and orientation of magnetic field. We admit that the best way 
to compare the accretion model to Faraday rotation measure is not by comparing 
accretion rates, but by self consistent calculations of RM from the model.

Profiles of temperature and density resulting from our fitting put constraints 
on the physical conditions in the vicinity of the GC. 
It may provide information on the value of the gas pressure of the hot plasma. 
Together with radiation pressure taken from observed luminosity states of Sgr~A* 
we can estimate the size of the thermal instability and eventual cloud formation 
in this region \citep{rozanska14}. Such multi-phase complex structure is observed in our GC 
\citep[for review, see][]{cox2005}. This structure is important from the point of
view of feeding the central black hole as well as the possibility of an
accompanying outflow. 

To study any possible outflow related tot he Sgr~A* accretion, 
we should search for spatially resolved regions of the hot plasma farther away 
from the Sgr~A* region, where the  line emission is prominent. 
For example, strong  iron K$_{\alpha}$ line emission detected in our 
observations occurs in the Sgr~A~East region.
The origin of this emission is strongly affected by SNR, and more advance spectral model 
should be used to describe the emerging spectra. We address these
studies in the forthcoming paper.  

%
\begin{acknowledgements}
We thank Roman Shcherbakov for providing {\it Chandra} ACIS-I effective 
area responce function.
This research was supported by  Polish National Science 
Center grants No. 2011/03/B/ST9/03281, 2013/10/M/ST9/00729, 
and by Ministry of Science and Higher Education grant W30/7.PR/2013. 
It has received funding from the European 
Union Seventh Framework Programme (FP7/2007-2013) under grant agreement No.312789. 
\end{acknowledgements} 

\bibliographystyle{aa}
\bibliography{refs}

\end{document}